\documentstyle[prl,aps,twocolumn,graphicx]{revtex}

\newcommand{\ud}{\,{\mathrm d}}
\newcommand{\zc}{\overline{z}}

\newcommand{\fc}{\overline{f}}

\newcommand{\uiint}{\int\!\!\!\int}
\newcommand{\uRe}{{\mathrm{Re}}\,}
\newcommand{\uIm}{{\mathrm{Im}}\,}
 
\begin{document}
\draft 
\title{Two-dimensional topological solitons in soft ferromagnetic
 cylinders} 
\author{Konstantin L. Metlov}
\address{Institute of Physics ASCR, Na Slovance 2, Prague 8, CZ-18221}
\date{\today} \maketitle
 
\begin{abstract}
  A simple approach allowing to construct closed-form analytical
  zero-field magnetization distributions in cylindrical particles of a
  small thickness and an arbitrary shape (not necessarily circular) is
  presented. The approach is based on reduction of the non-linear
  Euler equations for magnetization vector field to the classical
  linear Riemann-Hilbert problem. The result contains all the
  distributions minimizing the exchange energy functional and the
  surface magnetostatic contribution exactly, except for the
  neighbourhood of topological singularities on the cylinder faces
  where the result is approximate. The completeness of the analysis
  permitted to find a new type of a topological soliton in the case of
  circular cylinder.  Also, an example of magnetic vortex in a
  triangular cylinder is given to investigate the role of the particle
  corners.
\end{abstract}

Small magnetic particles shaped as flat arbitrary cylinders with the
size of the order of $100$ nanometers and their arrays recently
captured a lot of attention due to their unusual magnetic properties
(as unique systems where two-dimensional topological solitons can be
directly observed) and also because of their applications in the
magnetic random access memory devices.

In large ferromagnetic bodies the equilibrium magnetization
distributions are usually represented as a set of magnetic domains
separated by possibly bent one-dimensional domain
walls\cite{Hubert_Shafer}. Unfortunately, this approach is hard to
apply to small magnetic particles if their size is such that the
domain wall does not fit inside, resulting in considerable distortions
of its profile with respect to the one-dimensional case.
Consequently, it is necessary to consider this distribution as a
whole. A number of finite-element simulations of the problem were
already performed, see e.g. \cite{HK99} and references therein.

In this Letter an analytical approach allowing to find closed-form
expressions for equilibrium topologically charged magnetization
distributions in soft magnetic particles shaped as flat arbitrary
cylinders is presented.

\paragraph*{Mathematical formulation of the problem.}  Consider a
particle made of an isotropic magnetic material, the anisotropy of
soft material is neglected here. In the case of no applied magnetic
field its total energy has two contributions, the exchange and the
dipolar (magnetostatic). Let us limit ourselves to the cylindrical
particles with thickness $L<L_E=\sqrt{C/M_S^2}$ (it usually can be
relaxed\cite{UP93} for $L$ up to several $L_E$), where $C$ is the
exchange constant and $M_S$ is the saturation magnetization of the
material. In this case the magnetization distribution
$\vec{m}(\vec{r})=\{m_X,m_Y,m_Z\}=\vec{M}(\vec{r})/M_S$,
$|\vec{m}(\vec{r})|=1$ can be assumed uniform along the cylinder
thickness ($Z$ axis), with $\vec{r}=\{X,Y\}$ the radius vector in the
cylinder plane. In this notation the total energy of the particle (a
functional of $\vec{m}$) can be written in the following way
\begin{eqnarray}
  \label{eq:energy_2d}
  \frac{e[\vec{m}]}{M_S^2} = \!\! \uiint_{\cal D} 
  \left\{
    \frac{L_E^2 L}{2}\!\!\!\! \sum_{i=X,Y,Z} \!\!\!
    (\vec{\nabla} m_i)^2 - \!\!\!\!\!
    \int\limits_{-L/2}^{L/2}\!\!\! \vec{h}_D[\vec{m}]\cdot\vec{m}  \ud z  
  \right\}\ud^2 \vec{r},
\end{eqnarray}
where ${\cal D}$ is the area of the cylinder face,
$\vec{\nabla}=\{\partial/\partial X,\partial/\partial
Y,\partial/\partial Z\}$ is the gradient operator,
$\vec{h_D}[\vec{m}]=\vec{H}_D[\vec{m}]/M_S$ is the demagnetizing field
(a functional of $\vec{m}$) created by the magnetization distribution.
The demagnetizing field can be expressed using the Maxwell equations
through its scalar potential $\vec{H}_D=\vec{\nabla}U(\vec{r})$, which
is, in turn, a solution of the Poisson equation $\vec{\nabla}^2 U =
\rho$ with the requirement (due to the finite size of the particle)
that both $|\vec{r}| U$ and $|\vec{r}|^2 |\nabla U|$ are finite as
$|\vec{r}| \rightarrow \infty$, $\rho=\vec{\nabla}\cdot\vec{M}$ is the
density of magnetic charges (on the boundary equal to the normal
component of $\vec{M}$), \cite{Aharoni_book}.

The exact solution for equilibrium $\vec{m}(\vec{r})$ corresponds to
the minimum of the functional (\ref{eq:energy_2d}) and is given by a
system of non-linear integral partial differential equations. They are
hard to handle even with the help of computer for a particular case.
Thus, a simplification of the original problem (\ref{eq:energy_2d})
needs to be applied.

To make such simplification let us qualitatively analyze contributions
of the different energy terms in (\ref{eq:energy_2d}).  It turns out
that for sufficiently small particle there is a well defined hierarchy
of energies. The most important term is the exchange, as when the
particle gets smaller not only the amount of possible magnetic charges
decreases (on par with the exchange energy) but also (due to absence
of magnetic monopoles) the positive self energy of charges gets more
and more compensated by their negative interaction energy. It is also
clear, as the role of surface increases for small particle sizes, that
the surface magnetostatic contribution is more important then the
volume one. The surface contribution can be subdivided into the energy
of the magnetic charges on the faces and on the side of the cylinder.
Because for flat cylinders the faces have usually larger area than the
side, the former energy is more important.  Thus, in flat cylinders we
have the following energy hierarchy: exchange, face charges, side
charges, volume charges. Note that the exact criteria for the
necessary smallness of the cylinder strongly depend on its shape (due
to the long-range character of the dipolar interactions) and should be
analyzed on a case-by-case basis.  Let us just assume for now that the
cylinder satisfies the criteria.

The solution for the problem exactly minimizing the exchange energy
functional (\ref{eq:energy_2d}) with $\vec{h_D}=0$, and also having
negligible amount of the face magnetic charges present only at the points
of topological singularities was given in \cite{M01_solitons} as
\begin{equation}
  \label{eq:sol_SM}
  w(z,\overline{z})=\left\{
    \begin{array}{ll}
      f(z) & |f(z)| \leq 1 \\
      f(z)/\sqrt{f(z) \fc(\zc)} & |f(z)|>1
    \end{array}
    \right. 
\end{equation}
where $f(z)$ is an arbitrary analytic (in the sense of Cauchy-Riemann
conditions \cite{Lavrentiev_Shabat}) function of a complex variable,
line over a symbol denotes complex conjugation, straight brackets
denote absolute value of the complex number, $z=X+\imath Y$ is the
complex coordinate, $\imath=\sqrt{-1}$. The components of the
magnetization vector are given by $m_X+\imath m_Y = 2 w/(1+|w|^2)$ and
$m_Z=(1-|w|^2)/(1+|w|^2)$. This solution consists of two parts both
locally extremal to the exchange energy functional: the one
conventionally termed as soliton\cite{BP75} at $|f(z)|<1$ and the
meron\cite{G78}.  The role of the soliton ``hat'' ($|f(z)|<1$) is to
cover the topological singularity of the meron \cite{M01_solitons},
this is the only region of (\ref{eq:sol_SM}) where $|m_Z|>0$. Pure
solitons are not realized in considered particles due to excessive
face magnetic charges \cite{M01_solitons}.

Let us now see if using the freedom of choice for $f(z)$ in
(\ref{eq:sol_SM}) we can exactly remove the side magnetic charges
(whose energy is next in the hierarchy). This problem can be
formulated as: {\em to find a function $f(z)$ analytical in the region
  ${\cal D}$ in such a way that $\uRe [ f(\zeta) \overline{n(\zeta)}]
  = 0$ (no magnetization components normal to the side), where $\zeta
  \in {\cal C}=\partial {\cal D}$ is the boundary of ${\cal D}$, and
  $n(\zeta)=n_x (\zeta) +\imath n_y (\zeta)$ is the complex normal to
  ${\cal C}$}. Which makes it reduced to the well known linear
Riemann-Hilbert problem\cite{Lavrentiev_Shabat}.
  
To write the solution let us denote the conformal transformation of
the interior of the unit circle $|t|\leq 1$ to $z \in \cal D$ as
$z=T(t)$, according to Riemann theorem such transformation exists for
any simply connected region $\cal D$. Under such transformation $\cal
C$ transforms onto the boundary of the unit circle $|\lambda|=1$ and
the normal is given by $n_T(\lambda)=n(T(\lambda))$. Then, all the
solutions for $f(t)$ are given by
\begin{eqnarray}
  \label{eq:HP_sols}
  f(t) & = & \frac{a_0 t^2 + a_1 t + a_2}{e^{F^{+}(t)}} + 
  \frac{\overline{a_0} + \overline{a_1} t + \overline{a_2} t^2}
  {e^{\overline{F^{-}(1/\overline{t})}}} \\
  \label{eq:HP_sols_Cauchy}
  F(t) & = & \frac{1}{2\pi\imath}\oint_{|\lambda|=1} 
  \frac{\log[- \lambda^2 \overline{n_T(\lambda)}/n_T(\lambda)]}{\lambda-t}
  \ud \lambda,
\end{eqnarray}
where $a_0$, $a_1$, $a_2$ are arbitrary complex constants,
$F^{+}(t)=F(t), |t|<1$ and $F^{-}(t)=F(t), |t|>1$ are the values of
the Cauchy-type integral (\ref{eq:HP_sols_Cauchy}) having
discontinuity at $|t|=1$. The presence of exactly three arbitrary
constants is due to the index\cite{index} $\mathrm{ind}
[-\overline{n_T}/n_T] = -2$. 

The derivative of the conformal transformation $T'(t) = G(t) e^{\imath
  \varphi(t)}$ gives the scaling (the real function $G(t)$) and the
rotation angle (the real function $\varphi(t)$) performed by a
transformation at a point $t$, and $\lambda$ (when $|\lambda|=1$), is
a normal to the unit circle. This allows to express (up to the real
multiplier) the normal to the face boundary as $n_T(\lambda) \sim
\lambda T'(\lambda)$. The real multiplier in $n_T(\lambda)$ is
irrelevant because only the combination $\overline{n_T}/n_T$ enters
(\ref{eq:HP_sols_Cauchy}).  Then, the integral
(\ref{eq:HP_sols_Cauchy}) gives $F^{+}(t)=-\imath \pi - \log T'(t)$
and $F^{-}(t)=- \log \overline{T'(1/\overline{t})}$, where the
identity $1/\overline{\lambda}=\lambda$ and the residue theorem were
used. Substititing these into (\ref{eq:HP_sols}) gives the final
answer
\begin{equation}
  \label{eq:HP_sols1}
  f(t) = (i t c + A - \overline{A} t^2) T'(t),
\end{equation}
where $c=\uIm a_1$ is a real constant and $A=\overline{a_0}-a_2$ is a
complex constant, the role of these constants is best illustrated by
the following example.

\begin{figure}[htbp]
  \begin{center}
    \includegraphics[scale=0.5]{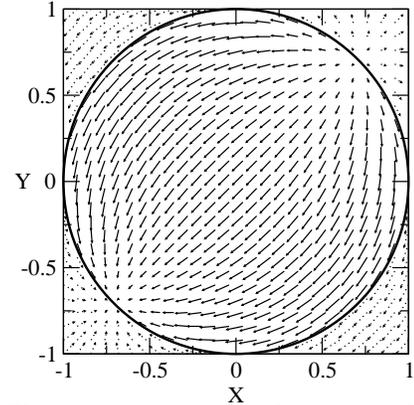}
    \caption{The magnetization distribution (in-plane components)
      within the circular cylinder having two skyrmions bound on the
      side (\ref{eq:HP_sols_circular}), $c=0$, $A=e^{\imath \pi/4}/2$.
      The dotted arrows show the solution continued into the region
      with no magnetic material to confirm that the magnetization
      non-uniformities are indeed skyrmions.}
    \label{fig:circ_cyl}
  \end{center}
\end{figure}

\paragraph*{Circular cylinder.} In this case
$T(t)=t$ and the magnetization distribution is (\ref{eq:sol_SM}) with
\begin{equation}
  \label{eq:HP_sols_circular}
  f_{\mathrm CI}(z) = i z c + A - \overline{A} z^2.
\end{equation}
The values of $c$ and $A$ need to be determined from the detailed
analysis of the energy including magnetostatic and can not be
determined based on the pole avoidance principle alone. However, the
interesting point is that apart of the well known vortex
solutions\cite{UP93} corresponding to $|c| > 2 |A|$ there are also
solutions of another type with $|c| < 2 |A|$ corresponding to a pair
of skyrmions\cite{S58} (also known as hedgehogs) bound to the cylinder
sides, see Fig.~\ref{fig:circ_cyl}.  These solutions are relevant to
the consideration of quasi-uniform magnetization states (by allowing
the pair of skyrmions to move away from the cylinder boundary) and of
a vortex nucleation. The nucleation of a vortex can be described as
symmetrical moving of the pair of skyrmions from infinity (uniform
state) to the particle boundary with quasi-uniform states forming in
the process and, then, by changing $\infty>|A|>c/2$ alongside the
particle boundary to the point where two skyrmions annihilate and the
vortex forms, moving subsequently to the particle center as $A$
changes down to $c/2>|A|>0$. This process will be considered in detail
in a forthcoming paper. Let us now see what happens if a particle has
corners by considering the next example.

\begin{figure}[htbp]
  \begin{center}
    \includegraphics[scale=0.5]{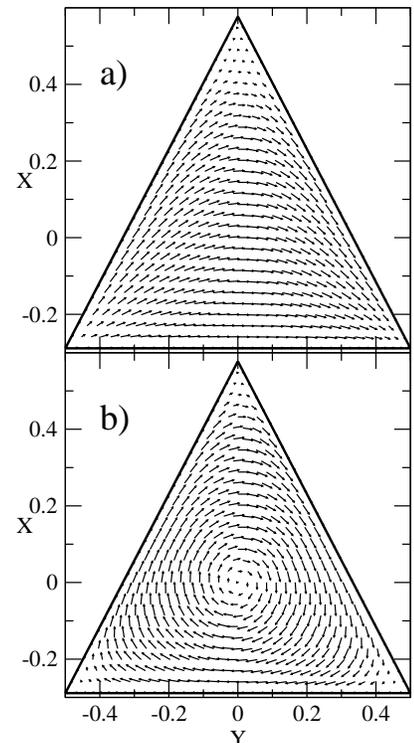}
    \caption{Magnetization distributions within triangular cylinder
      (\ref{eq:HP_sols_triangular}) : a) with two side-bound skyrmions
      $c=0$, $A=-2\imath$, $c_2=4$; b) with a vortex $c=1$, $A=0$, $c_2=4$.}
    \label{fig:triang_cyl}
  \end{center}
\end{figure}

\paragraph*{Triangular cylinder} In this case the conformal map is given
by the Schwartz-Christoffel integral $T(t)=C \int_0^t (1-t^3)^{-2/3}
\ud t$. The side length of the triangle is $1$ if
$C=2^{2/3}\sqrt{3\pi}/\Gamma(1/6)/\Gamma(1/3)$, $\Gamma(x)$ is Euler's
gamma function. Using (\ref{eq:HP_sols1}) we get
\begin{equation}
  \label{eq:HP_sols_triangular}
  f_{\mathrm TR}(z)= (i t c + A - \overline{A} t^2)
  \frac{1}{(1-t^3)^{2/3}}, \,\, t\rightarrow T^{-1}(z),
\end{equation}
where $T^{-1}(z)$ means inverse function of $T(t)$ which is unique
because the transformation is conformal. This expression diverges at
the corners of the triangle because corners represent topological
singularities of the type $\infty/\infty$ in the meron. To avoid them
we shall cover corners by a soliton ``hats'' similarly as it was
done\cite{M01_solitons} for the singularities of the type $0/0$
corresponding to the vortex centers.  Thus, instead of
(\ref{eq:sol_SM}) we shall use
\begin{equation}
  \label{eq:sol_SM2}
  w(z,\overline{z})=\left\{
    \begin{array}{ll}
      f(z) & |f(z)| \leq 1 \\
      f(z)/\sqrt{f(z) \fc(\zc)} & 1<|f(z)| \leq c_2\\
      f(z)/c_2 & |f(z)| > c_2,
    \end{array}
    \right. 
\end{equation}
where $c_2>1$ is an additional arbitrary real constant. The result is
plotted in Fig.~\ref{fig:triang_cyl}.

\paragraph*{Summary.} A general method is provided for obtaining
closed-form analytical expressions for magnetization distributions in
flat ferromagnetic cylinders of arbitrary shapes made of isotropic
material. This problem is shown to be equivalent to finding the
conformal transformation of the unit circle to the shape of the
cylinder face $T(t)$. Then, the magnetization distributions are given
by (\ref{eq:HP_sols1}) and (\ref{eq:sol_SM}) (or (\ref{eq:sol_SM2}) if
the particle has corners). No other magnetization distributions
exactly minimizing the exchange energy of the particle and having no
surface magnetic charges on its side exist.  A new type of soliton
solution was found by applying this procedure to the circular
cylinder, see Fig.~\ref{fig:circ_cyl} (and also experimental Fig.~1(c)
in \cite{PSN00}, however interpreted differently there).

The only term not included into this consideration is the
magnetostatic energy of the volume charges but, as it was
qualitatively argued, the effect of this term is the smallest for
small particles and is expected to introduce only minor corrections to
the magnetization distributions obtained here. If desired, these
corrections can be found by perturbative expansions or by numerical
analysis.

This work was supported in part by the Grant Agency of the Czech
Republic under projects 202/99/P052 and 101/99/1662. I would like to
thank Ivan Tom{\'a}{\v s}, and Konstantin Guslienko for reading
the manustcript and many valuable discussions.

\begin{thebibliography}{10}
\bibitem{index}
index of a complex function ${\mathrm ind} f$ on a contour $\partial {\cal D}$,
  is the number of full rotations of its argument when going around $\partial
  {\cal D}$ once positively, ${\mathrm ind} \, a/b ={\mathrm ind}\, a -
  {\mathrm ind}\, b$, ${\mathrm ind}\, \overline{a} = - {\mathrm ind}\, a$, and
  the index of a complex normal to $\partial {\cal D}$ is $1$.
\end{thebibliography}
\end{document}